\pdfoutput=1

\documentclass[aps,prl,twocolumn,groupedaddress,showpacs,showkeys]{revtex4-1}

\usepackage{natbib}
\usepackage{graphicx}
\usepackage{color}
\usepackage{soul}

\begin{document}

%Title of paper
\title{Cryogenic scanning force microscopy of quantum Hall samples:\\ Adiabatic transport originating  in anisotropic depletion at contact interfaces}
% version 10: \title{Cryogenics scanning force microscopy of the adiabatic transport \\ showing its origins to anisotropic contact interfaces}
% \title{Cryogenic scanning force microscopy of adiabatic transport in quantum Hall samples showing its origins in anisotropic contact interfaces}
%
%Authors
\author{F.~Dahlem}
\email[Present address: N\'eel Institute CNRS/UJF, Grenoble (France); ]{franck.dahlem@grenoble.cnrs.fr.}
\author{E.~Ahlswede}
\email[Present address: Zentrum f\"ur Sonnenergie- und Wasserstoff-Forschung Baden-W\"urtemberg, Industriestr.~6, D-70565 Stuttgart (Germany). ]{} 
\author{J.~Weis}
\author{K.~v.~Klitzing}
\affiliation{Max-Planck-Institut f\"ur Festk\"orperforschung, Heisenbergstr.~1, D-70569 Stuttgart, Germany }
\date{\today}
%
%
%%%%%%%%%%%%%%%%%%%%%%%%%%%%%%%%%%%%%%%%%%%%%%%%%%%%%%%%%%%%%%%%%%%%%%%%%%%
\begin{abstract}
Anisotropic magneto resistances and intrinsic adiabatic transport features 
are generated on quantum Hall samples based on an (Al,Ga)As/GaAs heterostructure with alloyed Au/Ge/Ni contacts.
We succeed to probe the microscopic origin of these transport features with a cryogenic scanning force microscope (SFM) by measuring the local potential distribution within the two-dimensional electron system (2DES). These local measurements reveal the presence of an incompressible strip in front of contacts with insulating properties depending on the orientation of the contact/2DES interface line relatively to the crystal axes of the heterostructure.  Such an observation gives another microscopic meaning to the term 'non-ideal contact' used in context with the Landauer-B\"uttiker formalism applied to the quantum Hall effect. 
\end{abstract}
\maketitle
%
%
%%%%%%%%%%%%%%%%%%%%%%%%%%%%%%%%%%%%%%%%%%%%%%%%%%%%%%%%%%%%%%%%%%%%%%%%%%%
%%%%%%%%%%%%%%%%%%%%%%%%%%%%%%%%%%%%%%%%%%%%%%%%%%%%%%%%%%%%%%%%%%%%%%%%%%%

% ------- Intro ------------------------
For now more than one decade, cryogenic scanning force microscopy (SFM) has been used to probe the electrostatic potential 
distribution in two-dimensional electron systems (2DES) realized in (Al,Ga)As/GaAs heterostructures 
under quantum Hall conditions~\cite{mccornick4654,weitz247,*ahlswede562,*ahlswede165}. Those measurements have demonstrated 
the important role of compressible and incompressible strips~\cite{chklovskii+shklovskii+glazman4026,*lier+gerhardts7757,*siddiki+gerhardts125315,*siddiki+gerhardts195335} for the current distribution in quantum Hall samples. 
Despite of these experimental findings,
the edge state picture~\cite{buttiker9375} remains the widely used model to describe the magneto-transport in quantum Hall devices, also in the context of topological insulators~\cite{roth294,*buttiker278}. 
Its success has been partly legitimated by its ability to include contact effects~\footnote{In controlled experiments, 'ideal and non-ideal' contact properties are mimiced by locally depleting the 2DES by gate electrodes in front of contacts.} and to explain adiabatic magneto transport features~\cite{haug131,*vonHouten+Beenaker1} --  like the disappearance of peaks in the Shubnikov-de Haas oscillations, the extension of quantum
Hall plateaus to lower magnetic fields and the existence of non-local resistances~\cite{komiyama+nii7,*alphenaar677,*mceuen2062,*wees8066}. 
In this letter we manage a comparison between magneto transport and scanning force microscopy investigations on Hall bar samples showing adiabatic transport features without the use of gates. 
The SFM measurements clearly present potential distributions from which we conclude that the incompressible strip in front of alloyed contacts
possesses different insulating properties depending of the orientation relatively to the crystal axes of the underlying heterostructure.
This result shows how to interpret adiabatic features in terms of compressible/incompressible strips and what is the microscopic origin of anisotropy in magneto resistances.

%%%%%%%%%%%%%%%%%%%%%%%%%%%%%%%%%%%%%%%%%%%%%%%%%%%%%%%%%%%%%%%%%%%%%%%%%%% 
\begin{figure}[b!]
$$\includegraphics[width=\columnwidth]{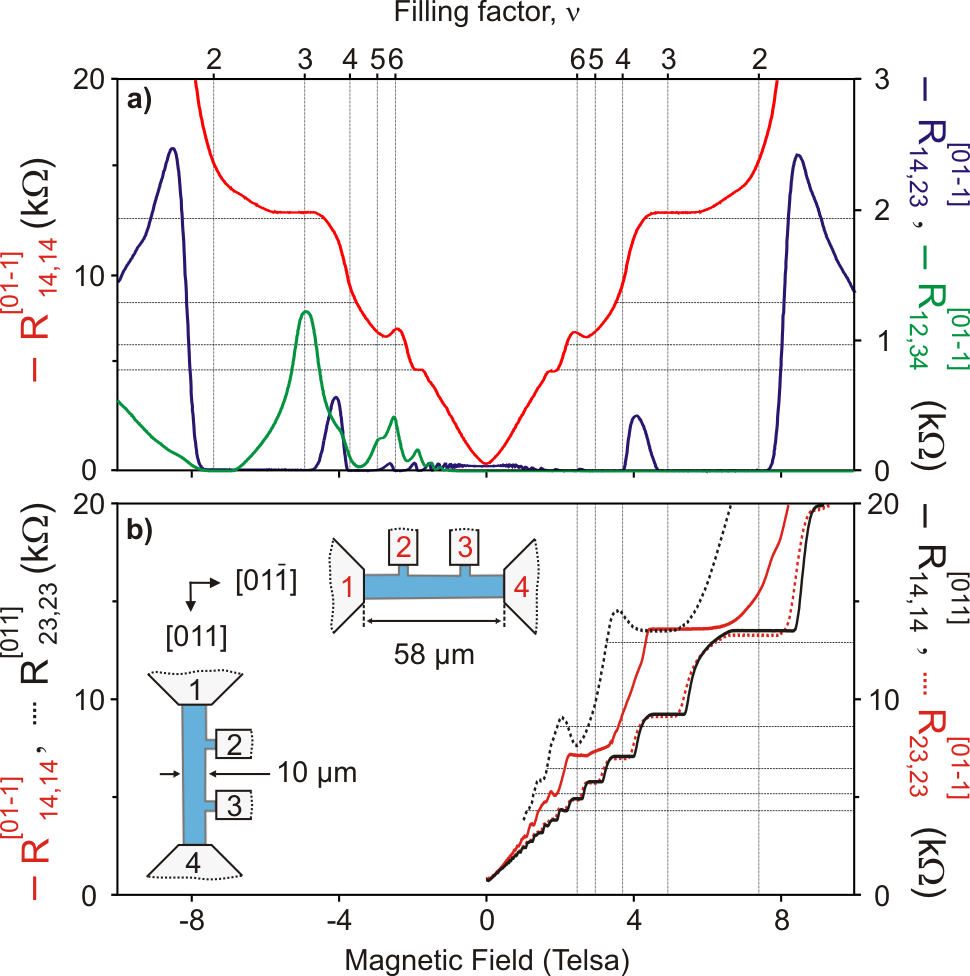}$$
\caption{Magneto resistances $R_{ij,kl}$ measured on the four-terminal Hall bar geometry shown in the inset: 
a) Hall plateaus in $R^{\rm [01-1]}_{14,14}$ (red curve)  are shifted to lower magnetic field values, Shubnikov-de Haas peaks in the longitudinal resistance $R^{\rm [01-1]}_{14,23}$ 
(blue curve) are missing between $\nu$ = 2 to 3, 4 to 6, and a non-local resistance $R^{\rm [01-1]}_{12,34}$ 
(green curve) is observed at negative magnetic field ($T=1.3$~K).
 b) Comparison of $R_{14,14}$ and $R_{23,23}$ measured on Hall bars oriented either along [011] or [01-1] direction ($T=26$~mK).
\label{Transport_6and4contact}}
\end{figure}
%%%%%%%%%%%%%%%%%%%%%%%%%%%%%%%%%%%%%%%%%%%%%%%%%%%%%%%%%%%%%%%%%%%%%%%%%%%
%-------The samples ------------------------------------------------------------------------------------------------------------------
The samples used here are based on an (Al,Ga)As/GaAs heterostructure containing the 2DES at the heterojunction interface 60~nm below the surface.
The electron density and mobility are respectively $\rm{3.6\times 10^{15}~m^{-2}}$ and $\rm{160~m^{2}(Vs)^{-1}}$ at $1.3$~K. These conditions lead to $l\approx\rm{13~\mu m}$ for the electron mean free path, which is three time larger than those of the samples measured in~\cite{weitz247,*ahlswede562,*ahlswede165}. High electron mobility Hall bars as depicted in Fig.~\ref{Transport_6and4contact} are defined by optical lithography and wet etching.  
Following our standard recipe~\cite{goktas1579}, Au/Ge/Ni film is alloyed to achieve low resistive contacts to the 2DES. 
After annealing, the remaining 2DES has a size of about 10~$\rm{ \mu m}$ width and 58~$\rm{\mu m}$ length, which is comparable to the scanning range 20~$\rm{\mu m}$ by 20~$\rm{\mu m}$ of our cryogenic SFM.
From transmission line measurements~\cite{goktas1579}, we also find on these samples an anisotropy in contacting the 2DES in [01-1] or in [011]  
crystal direction which is a general property of alloyed Au/Ge/Ni contacts~\cite{kamada1263, goktas1579}. Here the specific resistivity $r_{\rm c}\equiv R_{\rm c}\cdot w$ (w, the width of the borderline between 2DES and alloyed metal) is $r_{\rm c}\approx \rm{0.25~\Omega mm}$ for contacts with their borderline perpendicular to [011], and $r_{\rm c}\approx ~\rm{0.45~\Omega mm}$ for contacts with their borderline perpendicular to [01-1].

%--------- Transport Data -----------------------------------------------------------------------------------------------------------------
To obtain the magneto resistance $R^{\rm [01-1]}_{ij,kl}$ (Hall bar oriented along [01-1]), an  ac  current of 100~nA rms amplitude is biased via contacts $i$ and $j$, and the ac voltage drop between contact $k$ and $l$ is measured by lock-in technique. In comparison to a standard six-terminal Hall bar of same size, the four-terminal Hall bars with two contacts on one  side of the mesa display in  some magneto resistances $R^{\rm [01-1]}_{ij,kl}$ generic adiabatic transport features, reproduced on many samples of same geometry. Examples are given in  Fig.~\ref{Transport_6and4contact}a:
(1) In measuring the four-terminal longitudinal resistance $R^{\rm [01-1]}_{14,23}$, resistance peaks of the Shubnikov-de Haas (SdH)  oscillations are completely suppressed in between $\nu = 3$ and  $2$,  and in between $\nu = 6$ and  $4$~\footnote{The filling factor is obtained from the electron density $n_{\rm s}$ and magnetic field  $B$ by $\nu=2\pi\hbar\,n_{\rm s}/(e\,B)$ .}.  
(2) The non-local resistance
$R^{\rm [01-1]}_{12,34}$ asymmetrically changes with the magnetic field orientation, i.e. for positive magnetic field
no voltage drop is observed between contact 3 and 4 ($R^{\rm [01-1]}_{12,34}=0$), while for negative sign non-zero resistance values are measured. 
(3) For the two-terminal resistance $R^{\rm [01-1]}_{14,14}$, Hall plateaus with $R=h/(i\,e^2)$ are shifted to lower magnetic field values corresponding to filling factors around $\nu=i+1$. In general we found that if one of two contacts 
has its  interface line with the 2DES perpendicular to [01-1], Hall plateaus in the two-terminal measurements ($i=k$, $j=l$) are shifted to lower magnetic field values.
The effect becomes most striking when directly comparing $R_{14,14}$ and $R_{23,23}$ of two Hall bars of same geometry but either oriented along [011] or  [01-1]
crystal direction (see Fig.~\ref{Transport_6and4contact}b): The two-terminal Hall resistances $R^{\rm [011]}_{14,14}$ and $R^{\rm [01-1]}_{23,23}$ superpose whereas $R^{\rm [01-1]}_{14,14}$ and $R^{\rm [011]}_{23,23}$ deviate.
All these deviations (1) to (3) in the magneto resistances have been described in literature and are usually interpreted within the edge-state picture as a non-equilibrium situation between edge states running along the same edge, caused by full or partial reflection of edge states at imperfect contacts \cite{haug131}. 
The contacts here are however state-of-the-art for both orientations, i.e. good ohmic and low resistive. 
Furthermore, we succeed to generate intrinsic adiabatic transport features by simply selecting the mesa orientation and its geometry~\cite{oswald235331}. 

%------ Scanning Probe ----------------------------------------------------------------------------------------------------------------------
%%%%%%%%%%%%%%%%%%%%%%%%%%%%%%%%%%%%%%%%%%%%%%%%%%%%%%%%%%%%%%%%%%%%%%%%%%%
\begin{figure}[b]
$$\includegraphics[width=\columnwidth]{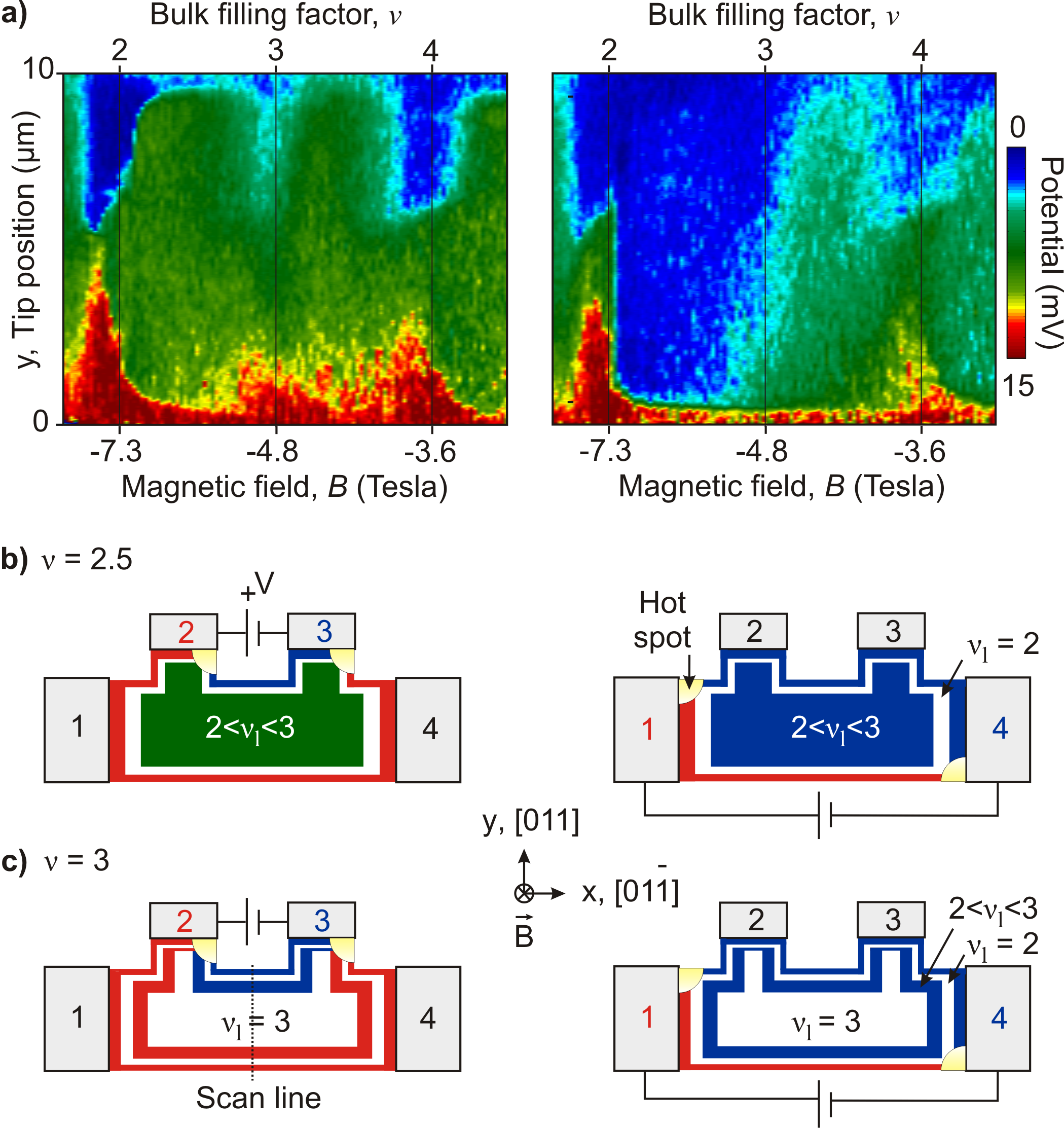}$$
\caption{(a) Hall potential profiles at negative magnetic fields measured with a cryogenic scanning force microscope. 
The local potential is recorded along the 10 $\rm \mu$m width in the center of the mesa for measurement configurations 
$R^{\rm [01-1]}_{23,kl}$ (left) and $R^{\rm [01-1]}_{14,kl}$ (right). 
(b),(c) Sketch of the (in-)compressible regions and  potential landscape for $\nu=2.5$  and $\nu=3$, respectively.
Further incompressible strips at the edges - if anyhow existing - are suppressed.
\label{non_ideal}}
\end{figure}
%%%%%%%%%%%%%%%%%%%%%%%%%%%%%%%%%%%%%%%%%%%%%%%%%%%%%%%%%%%%%%%%%%%%%%%%%%%
To find the microscopic origins of the previously described adiabatic transport features, the measurements of the Hall potential distribution within the 2DES have been performed by using a scanning force microscope operating at 1.3 K up to 13 Tesla~\cite{weitz349}. After finding the Hall bar structure at low temperature in an usual SFM
contact mode, the tip is retracted and made oscillating 70~nm above the mesa surface to probe only electrostatic forces. 
A constant bias voltage is applied between 2DES and metalized cantilever tip compensating for work function differences and avoiding a gating effect on the 2DES by the tip.
As the 2DES is buried in the heterostructure, a special calibration technique~\cite{weitz349} has to be used for extracting the Hall potential, 
i.e. the change of the local electrostatic voltage with current flow through the sample.
In a first scan, a $V_{\rm m}=$15~mV excitation at $w_{\rm m} =$3.4~Hz is applied to the whole 2DES and the modulation amplitude of the cantilever resonance frequency shift $\Delta f_{\rm res}^{w_{\rm m}}$ is recorded by lock-in technique along the mesa width. 
This signal -- being proportional to the local excitation, is almost constant with some local variations $M(y)$ due to the donor and surface charges embedded between 2DES and cantilever tip: $V^{(1)}(y)=M(y) \cdot V_{\rm m}$. 
A second scan is run at the same line with the potential excitation applied only to one contact whereas a second contact is grounded. 
A 3.4~Hz ac current flows through the sample and the related Hall potential distribution can be detected by a new $\Delta f_{\rm res}^{w_{\rm m}}$ measurement: $V^{(2)}(y)=M(y) \cdot V(y)$. 
The ratio between second and first scan $V^{(2)}(y)/V^{(1)}(y)$ delivers the normalized Hall potential profile $V(y)/V_{\rm m}$ with a spatial lateral resolution better than 100 nm~\cite{weitz349}.

In Fig.~\ref{non_ideal}a, the scanned potential profiles under the condition for measuring $R^{\rm [01-1]}_{23,kl}$  and $R^{\rm [01-1]}_{14,kl}$ are presented in a colorscale for negative magnetic field (i.e. clockwise rotation of electrons on  
cyclotron orbits) ranging from below $\nu=2$ to above $\nu=4$. The scans are taken in the middle of the 10~$\rm{\rm\mu m}$ wide mesa as indicated in the Hall bar sketches (Fig.~\ref{non_ideal}b,c). The colors blue, green and red in the plots correspond to the lowest, middle and highest potential value, respectively.
The profiles displayed in  Fig.~\ref{non_ideal}a(left) reflect the results also obtained in the past on samples with lower electron mobility of 50~$\rm{m^{2}(V\,s)^{-1}}$~\cite{ahlswede562}: The Hall potential drops on both sides of the Hall bar almost symmetrically at the expected positions of the innermost incompressible strips which merge around $\nu=i$ covering the whole bulk of the 2DES. At this latter filling factor the Hall potential drop is then widely spread in the bulk and $R^{\rm [01-1]}_{23,23}$, shown in Fig.~\ref{Transport_6and4contact}b, behaves normal. 

The situation is different in Fig.~\ref{non_ideal}a(right), where the change of potential happens completely on the lower side of the mesa~\cite{[{A similar profile was observed on a gated Hall bar by }]woodside238}. 
This behaviour is visible for the whole range between $\nu=2$ and $2.7$, and is still strongly pronounced above $\nu=2.7$. In this regime the bulk of the 2DES being compressible with $\nu_l>2$, the complete  Hall potential drop happens between the compressible bulk and the compressible edge over the incompressible strip $\nu_l=2$ at the lower mesa side [see Fig.~\ref{non_ideal}b(right)]. The current is thus driven without dissipation within this strip by the complete Hall voltage $V_{\rm H}$, $I=\nu_l\,e^2/h\,V_{\rm H}$~\footnote{The current is hence concentrated in a region of carrier density smaller than the bulk density as it was expected from the shift of Hall plateaus to lower magnetic fields.}.
As the longitudinal resistance $R^{[01-1]}_{14,23}$ vanishes at these filling factors and for both magnetic field signs, we can state that there is no equilibration between the compressible bulk with $2<\nu_l<3$ and contacts 1 and 4 [see Fig.~\ref{non_ideal}b(right)]. This decoupling comes from the depletion region in front of contacts 1 and 4 which have an enough smooth gradient towards the bulk to create a wide incompressible strip, which isolates the compressible edge and bulk under high voltage drops (For reminder, contacts with such orientation have a higher specific contact resistivity at zero magnetic field). Turning around the mesa orientation and measuring $R^{\rm [011]}_{14,14}$, the usual transport is obtained (see Fig.~\ref{Transport_6and4contact}b), i.e. the isolating property of the incompressible strip in front of contacts 1 and 4 with its interface line perpendicular to [011] is not pronounced at $2<\nu<3$ (Such contacts have a lower specific contact resistivity at zero magnetic field). To indicate the different isolating properties of incompressible strips in front of the contacts, their widths are marked smaller or wider in Fig.~\ref{non_ideal}b,c. 
At lower magnetic field, the two-terminal resistance $R^{[01-1]}_{14,14}$ shows a Hall plateau with $R=h/(2\,e^2)$ (see Fig.~\ref{Transport_6and4contact}a) even at $\nu=3$ where the bulk is mainly incompressible [see Fig.~\ref{non_ideal}c(right)]. It follows that the compressible strip with $2<\nu_l<3$ which encircles the incompressible bulk, stays isolated from the compressible edge with $\nu_l<2$ by the incompressible strip $\nu_l=2$. The potential drop is hence located between the two compressible strips at the lower mesa side, -- a situation which  reminds on the adiabatic transport model of non-equilibrium between edge states.

%%%%%%%%%%%%%%%%%%%%%%%%%%%%%%%%%%%%%%%%%%%%%%%%%%%%%%%%%%%%%%%%%%%%%%%%%%%
\begin{figure}[t]
$$\includegraphics[width=\columnwidth]{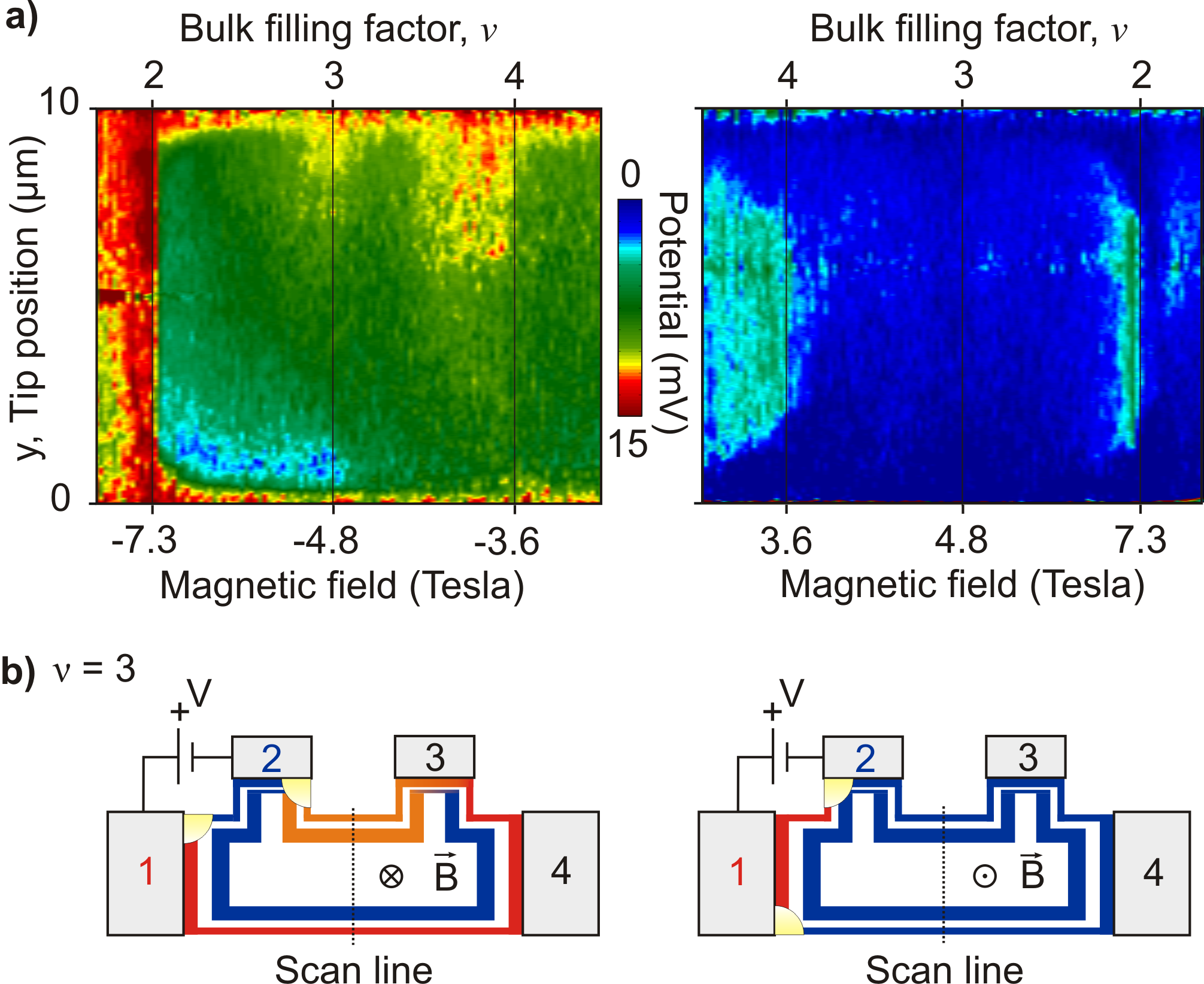}$$
\caption{(a) Local potential distribution probed for the measurement configuration $ R^{\rm [01-1]}_{12,kl}$. The potential profiles are different in switching the magnetic field orientation. At negative magnetic field in between $2<\nu<3$, a high potential (red) followed by a low potential (blue) is present at one edge.  
In contrary, the potential stays at low value (blue) for positive magnetic field. (b) Sketch of the (in-)compressible regions and  potential landscape for $\nu=3$.  \label{nonlocal}}
\end{figure}
\begin{figure}[t]
$$\includegraphics[width=\columnwidth]{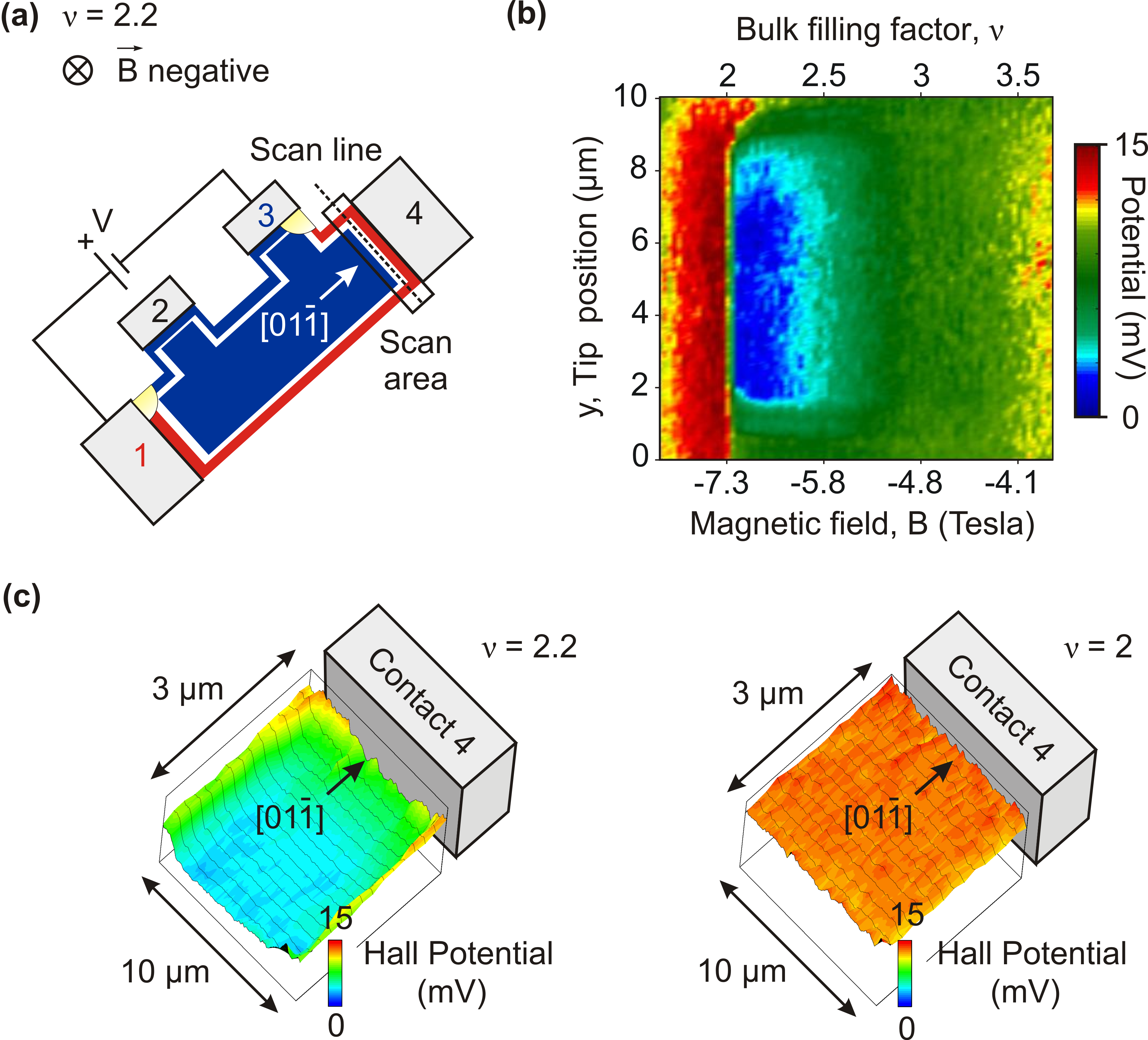}$$
\caption{(a) Sketch of Hall bar showing our (in)-compressible strip model in $R^{\rm [01-1]}_{13,kl}$ configuration. The positions of the SFM measurements are indicated. (b) Corresponding potential distribution for various negative magnetic fields probed on a scan line located few microns away from the voltage probing contact 4. (c) $x$-$y$ potential mapping measured $200$~nm away from the contact 4 at $\nu=2.2$ and $\nu=2$. At $\nu=2.2$ the potential remains high (red) along the mesa border whereas the inside has low potential (blue).
\label{voltagecontact}}
\end{figure}

%%%%%%%%%%%%%%%%%%%%%%%%%%%%%%%%%%%%%%%%%%%%%%%%%%%%%%%%%%%%%%%%%%%%%%%%%%%
In Fig.~\ref{nonlocal} the potential profiles correspond to the measurement configuration $R^{\rm [01-1]}_{12,kl}$.
For a negative magnetic field between $\nu=2$ and $\nu=3$, the potential drops at the upper mesa side from the edge to the bulk, remains constant over the bulk and drops towards the lower mesa side to increase again at the edge. This unusual potential profile is observed for almost the complete  magnetic field range. Obviously, the current driven between contact 1 and 2  runs along the whole edge of the sample passing by contact 3 and 4. Indeed, a non-local resistance  $R^{\rm [01-1]}_{12,34}$  has been found for negative magnetic field. For positive fields, no resistance is detected and consistently, the related potential distribution given in Fig.~\ref{nonlocal}b shows a constant potential from upper to lower mesa side, i.e. no current passing forth and back through this cross section.
As indicated in Fig.~\ref{nonlocal}b, a decoupling of the bulk and the contacts 1 and 4 is required to explain the presence of non-local resistances. 
Fig.~\ref{voltagecontact} presents the potential distribution in the vicinity of contact 4 for the measurement arrangement $R^{\rm [01-1]}_{13,kl}$. It directly proofs the decoupling by an incompressible strip at $\nu=2.2$: The potential of the compressible bulk is low even thus  the surrounding edge regions on the lower mesa side, in front of the contact 4 and at the upper mesa side are high in potential.
%%%%%%%%%%%%%%%%%%%%%%%%%%%%%%%%%%%%%%%%%%%%%%%%%%%%%%%%%%%%%%%%%%%%%%%%%%
%%%%%%%%%%%%%%%%%%%%%%%%%%%%%%%%%%%%%%%%%%%%%%%%%%%%%%%%%%%%%%%%%%%%%%%%%%%

%------ Summarize ----------------------------------------------------------------------------------------------------------------------
In summary, by comparing magneto transport and scanning probe investigations, we could identify that the presence of an incompressible strip in front of state-of-the-art ohmic contacts might cause an electrical decoupling of the compressible 2DES bulk region from the contacts. Important, we discover that the insulating properties of this incompressible strip depend on the orientation of the interface line 2DES/alloyed contact relatively to the underlying (Al,Ga)As/GaAs heterostructure. A contact with its interface line perpendicular to [011] equilibrates the compressible bulk and edge, whereas at the same time contacts with the interface line perpendicular to [01-1] does not. 
This anisotropy is related to different electron density profile at the contact interface which determines in high magnetic field the incompressible strip thickness. Adiabatic transport features hence originates in high electron mobility Hall bar with a geometry having contacts arranged on only three of four sides of the mesa.
We conclude, non-ideal (ideal) contact behaviour in terms of the Landauer-B\"uttiker formalism can be attributed to a smooth (sharp), partial  electrostatic depletion in front of ohmic contacts allowing for an large (thin) isolating incompressible strip.

%------ acknowledgments ----------------------------------------------------------------------------------------------------------------------
\begin{acknowledgments}
The authors thank M. Riek and T. Reindl for  
sample processing help, O.~G\"oktas and Y.~Boss for their
collaboration in study alloyed contact properties, and
A.~Siddiki, R.R.~Gerhardts,
D.~Quirion and A.C.~Welker for discussions. The 2DES wafer was provided
by W.~Wegschneider.
We acknowledge the Deutsche
Forschnungsgemeinschaft (DFG) for financial support under WE 1902/1.
\end{acknowledgments}

\bibliography{dahlem_biblio}

\end{document}